\documentclass[10pt,twocolumn,letterpaper]{article}

\usepackage{3dv}
\usepackage{times}
\usepackage{epsfig}
\usepackage{graphicx}
\usepackage{amsmath}
\usepackage{amssymb}
\usepackage{algorithm}
\usepackage{bbm}
\usepackage{caption}
\usepackage{subcaption}
\usepackage{hyperref}
\usepackage{fancyhdr}

\threedvfinalcopy % *** Uncomment this line for the final submission

\def\threedvPaperID{401} % *** Enter the 3DV Paper ID here

\ifthreedvfinal\fi
\begin{document}

%%%%%%%%% TITLE
\title{Ray Tracing-Guided Design of Plenoptic Cameras}

\author{Tim Michels and Reinhard Koch\\
	Kiel University, Germany\\
	{\tt\small \{tmi,rk\}@informatik.uni-kiel.de}
}

\maketitle
\thispagestyle{fancy}
\renewcommand{\headrulewidth}{0pt}
\fancyfoot[C]{\footnotesize{$\copyright$ 2021 IEEE. Personal use of this material is permitted.
		Permission from IEEE must be obtained for all other uses, in any current or future
		media, including reprinting/republishing this material for advertising or promotional
		purposes, creating new collective works, for resale or redistribution to servers or
		lists, or reuse of any copyrighted component of this work in other works.
		DOI: \href{https://doi.org/10.1109/3DV53792.2021.00120}{10.1109/3DV53792.2021.00120}}}

%%%%%%%%% ABSTRACT
\begin{abstract}
   The design of a plenoptic camera requires the combination of two dissimilar optical systems, namely a main lens and an array of microlenses. And while the construction process of a conventional camera is mainly concerned with focusing the image onto a single plane, in the case of plenoptic cameras there can be additional requirements such as a predefined depth of field or a desired range of disparities in neighboring microlens images. Due to this complexity, the manual creation of multiple plenoptic camera setups is often a time-consuming task.\\
   In this work we assume a simulation framework as well as the main lens data given and present a method to calculate the remaining aperture, sensor and microlens array parameters under different sets of constraints. Our ray tracing-based approach is shown to result in models outperforming their pendants generated with the commonly used paraxial approximations in terms of image quality, while still meeting the desired constraints. Both the implementation and evaluation setup including 30 plenoptic camera designs are made publicly available.
\end{abstract}

%%%%%%%%% BODY TEXT
\section{Introduction}

\begin{figure}[h]
	\centering
	\includegraphics[width=0.75\linewidth]{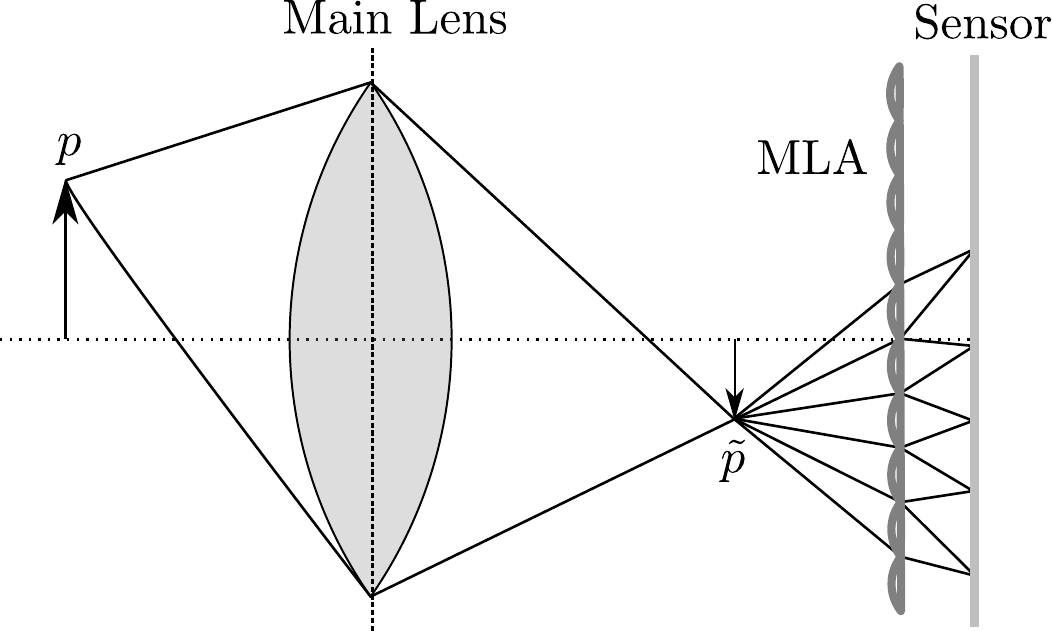}
	\caption{Schematic FPC: The virtual image $\tilde{p}$ of the scene point $p$ is imaged onto the sensor by multiple microlenses.}
	\label{fig_fpcsimple}
\end{figure}

Plenoptic cameras based on the ideas of Lippmann \cite{lippmann1908integralphoto} and Ives \cite{ives1928camera} can be regarded as conventional cameras with an additional microlens array (MLA) placed between the main lens and the sensor. This design enables applications such as post-shot refocusing and depth reconstruction from single images. During the past decades, two designs have been extensively studied and made commercially available. First, the standard plenoptic camera (SPC) \cite{adelsonwang1992plenopticcam}\cite{ng2005lightfieldcamera} has been proposed and its configuration simply requires the microlenses to be focused at infinity, i.e. the distance between MLA and sensor has to be the focal distance of the microlenses. Accordingly, all pixels behind a single microlens show approximately the same scene section from a slightly different perspective, resulting in a large angular resolution. Later the focused plenoptic camera (FPC) \cite{georgiev2006lightfieldcamdesign}\cite{perwass2012single} was proposed with the idea of using the microlenses to look at the virtual image of the scene instead of directly breaking it down into its directional components as in an SPC. This configuration, visualized in Fig. \ref{fig_fpcsimple}, proved to be beneficial in preserving a larger part of the conventional camera's spatial resolution at the cost of a certain loss of angular resolution compared to the SPC.
\\
Unfortunately, the construction of these cameras is not trivial since the focus matching between main lens and MLA is usually not restricted to a single plane in the scene, but instead the camera parameters need to be optimized to have a predefined depth of field (DoF). 
\begin{figure}[h]
	\centering
	\begin{subfigure}[b]{0.3\linewidth}
		\centering
		\includegraphics[width=\textwidth]{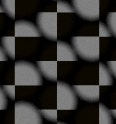}
	\end{subfigure}
	\hfill
	\begin{subfigure}[b]{0.3\linewidth}
		\centering
		\includegraphics[width=\textwidth]{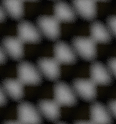}
	\end{subfigure}
	\hfill
	\begin{subfigure}[b]{0.3\linewidth}
		\centering
		\includegraphics[width=\textwidth]{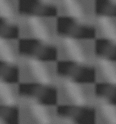}
	\end{subfigure}
	\caption{Desired image properties (left), defocus (middle) and heavily overlapping MLIs (right). }
	\label{fig_examples}
\end{figure}
Furthermore, the depth reconstruction often involves stereo matching between neighboring microlenses \cite{fleischmann2014plenoptic}. So it can also be beneficial to keep the disparities, i.e. the movement of a scene point's image from one microlens image (MLI) to the next one, within a range that can be handled well by the matching algorithm. And finally, there should be no large overlap of neighboring MLIs to prevent having pixels with ambiguous sources for the incoming light as shown in Fig. \ref{fig_examples}. 
All these additional constraints make a repeated manual reconfiguration of a self-build plenoptic camera a tedious and time-consuming task, while the alternative, the acquisition of multiple commercially available plenoptic cameras, is expensive. Therefore the creation of various setups with a realistic simulation can significantly reduce the time and money required to perform a meaningful evaluation of algorithms concerned with plenoptic camera data.\\
In this work we propose a method combining classical paraxial approximations, the lensmaker's equation and ray tracing using a full lens model in order to estimate the parameters of a plenoptic camera for a given set of constraints. Hereby, we assume a realistic, ray tracing-based simulation as well as main lens data given and use these models for the ray tracing part of our method to produce optimal parameters within the simulation framework. For the evaluation of our method we use the simulation presented in our previous work \cite{michels2018simulation}. We compare the models created with the thin and thick lens equation, as used in the related work, to our models with respect to the resolving capacity and find significant improvements. Furthermore, we also describe a method to calculate the DoF for off-center pixels to analyze the DoF across the sensor instead of only for the center pixel. In summary, our contributions are:
\begin{itemize} \itemsep0em
	\item A method for FPC parameter optimization outperforming the standard models
	\item A meta optimization procedure to achieve a predefined DoF
	\item A DoF calculation method for off-center pixels
	\item Publicly available\footnote{\href{https://gitlab.com/ungetym/fpc-design}{https://gitlab.com/ungetym/fpc-design}} implementations which can generate configurations directly importable by the simulation of \cite{michels2018simulation}.
	\item A dataset of 30 plenoptic camera designs
\end{itemize}
Here, we would like to note, that despite only describing the more complex case of design optimizations for FPCs, our method can easily be reduced to also optimize SPC parameters since these have fewer constraints due to the fixed positioning of the MLA exactly at the image side focus distance of the main lens, i.e. the distance to the main lens at which a sensor would be placed in a conventional camera.

\section{Related Work}
While the realistic simulation of conventional cameras via ray tracing with explicit lens models as in Kolb et al. \cite{kolb1995lensmodelforsimulation} and Wu et al. \cite{wu2011lensmodelforeffectsimulation} or with learned lens models as proposed in Zheng et al. \cite{zheng2017neurolens} have been extensively studied, the simulation of plenoptic cameras has often been based on oversimplified models. Using pinhole cameras or thin lens approximations for the main lens, the methods of  Fleischmann et al. \cite{fleischmann2014plenoptic}, Zhang et al. \cite{zhang2015forwardsimulation} and Liang et al. \cite{liang2015simuWOmainlens} are not able to accurately produce the same aberration effects as real plenoptic cameras. The more advanced approaches in Liu et al. \cite{liu2015numericalsimulation} and Li et al. \cite{li2017numericalsimulation} have been described theoretically, but not made publicly available. Only recently, with our previous work \cite{michels2018simulation} as well as Nürnberg et al. \cite{nurnberg2019simulation}, simulations of plenoptic cameras without considerable simplifications of the lens geometry became openly available. In this work we will use the former for the evaluation due to its ease of use concomitant with the integration in Blender \cite{blender}.\\
The situation regarding the automated design optimization of cameras is comparable. For conventional cameras, optimization methods have been used for decades and are already part of the standard literature \cite{kingslake2009lens}\cite{shannon2012applied} as well as industrial optical design tools like Zemax OpticStudio \cite{zemax}. Despite the small conceptual difference between a plenoptic camera and a standard camera of only one additional layer of lenses, the standard approaches of conventional lens design are not sufficient for the optimization of a plenoptic camera setup since the aforementioned additional constraints are not factored in. However, there is previous work explicitly concerned with optimizing certain parts of a plenoptic camera setup. For SPCs Ng et al. \cite{ng2005lightfieldcamera} describe the thin lens approximation based spacings between main lens, MLA and sensor and also propose the matching of main lens and microlens f-numbers in order to optimally cover the sensor area with MLIs. Hahne et al. \cite{hahne2014light} use paraxial ray tracing to analyze a given SPC's properties and provide a framework \cite{hahne2019plenoptisign} which can be used to quickly calculate these properties and hereby facilitating the manual design optimization. For FPCs, Perwass and Wietzke \cite{perwass2012single} describe the camera geometry in terms of the thin lens model and also provide a detailed scene side and image side DoF analysis along the optical axis.\\ 
In summary, there is only very limited work on the optimization of plenoptic camera parameters, and specifically FPCs, none of which features setting a predefined DoF and disparities or uses a full geometric model of the camera instead of paraxial approximations.

\section{Organization}
In section \ref{section_rtmeasure} we will discuss how to measure certain properties of a plenoptic camera based on ray tracing. Then, in section \ref{section_thinlens} and \ref{section_thicklens} the thin and thick lens based design considerations will be explained. Subsequently, we describe, how to use ray tracing to replace these approximations with more accurate measurements in section \ref{section_refinement}. To enable the usage of a predefined DoF, we will then describe a meta optimization procedure to match the camera's and the preset DoF in section \ref{section_dof_matching} and evaluate everything in section \ref{section_evaluation}.
\section{Ray Tracing-based Measurements}\label{section_rtmeasure}
Before describing the initial estimation and optimization of the FPC parameters, we will shortly discuss how to use ray tracing to measure some of the optical system's properties.
\subsection{Focus Calculation}\label{section_measure_focus}
\begin{figure}[h]
	\centering
	\includegraphics[width=0.8\linewidth]{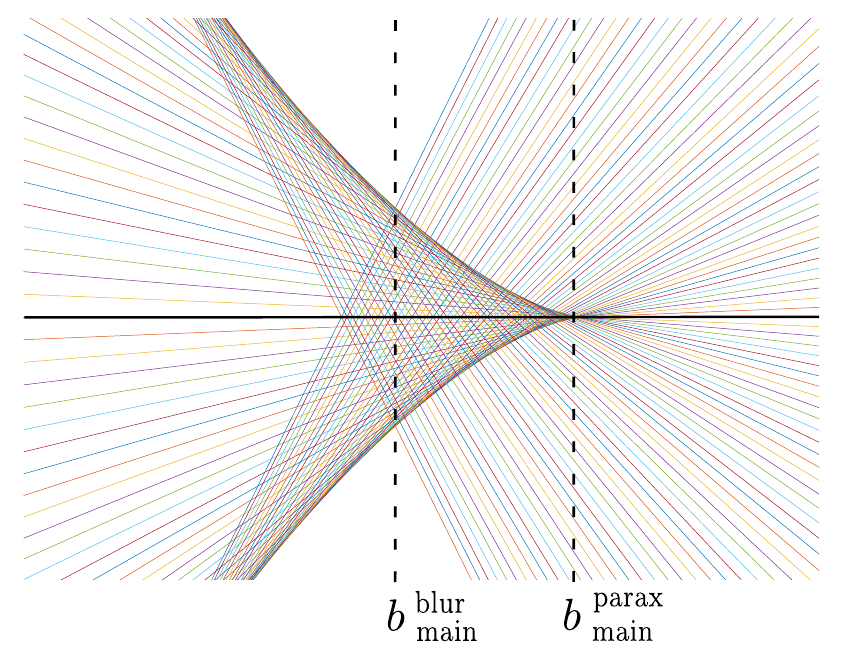}
	\caption{Visualization of different focus points.}
	\label{fig_focus}
\end{figure}

As explained by Geary \cite[chapter 33.5]{geary2002introduction}, there are several criteria for the placement of the image plane in a conventional camera. The resulting image plane positions are all located between the paraxial focus point $b_{\text{\hspace{1pt}main}}^{\text{\hspace{1pt}parax}}$ and the point of minimum blur $b_{\text{\hspace{1pt}main}}^{\text{\hspace{1pt}blur}}$, i.e. the distance from the main lens at which the envelope of traced rays has the minimum diameter as visualized in Fig. \ref{fig_focus}. 
The former can be calculated by tracing only a few rays from the scene point to focus on, close to the optical axis and calculating their intersections with this axis within the camera. On the other hand, the point of minimum blur can be calculated by tracing a bundle of rays through the currently investigated optical system with the real aperture value and searching for the minimum of the caustic curve, which for a bundle of rays is given as a piecewise linear function. To calculate a single focus point for the optical system, we observe, that the distances of the best image planes to the paraxial focus point according to \cite{geary2002introduction} are directly related by constants. More specifically, the distance between the point of minimum blur and the paraxial focus point and distance between the empirical best visual point and the paraxial focus point are described as
\begin{equation}
b_{\text{\hspace{1pt}main}}^{\text{\hspace{1pt}parax}} -b_{\text{\hspace{1pt}main}}^{\text{\hspace{1pt}blur}} = 8\cdot(f/\#)^2\cdot(-1.5\cdot W_{040})
\end{equation}
and
\begin{equation}
b_{\text{\hspace{1pt}main}}^{\text{\hspace{1pt}parax}} -b_{\text{\hspace{1pt}main}}^{\text{\hspace{1pt}BV}} \approx 8\cdot(f/\#)^2\cdot(-0.531\cdot W_{040})
\end{equation}
with $(f/\#)$ being the main lens f-number and $W_{040}$ denoting the wavefront aberration coefficient of the Seidel polynomial characterizing the lens aberrations (compare \cite[chapter 7.4]{geary2002introduction}). Since $b_{\text{\hspace{1pt}main}}^{\text{\hspace{1pt}blur}}$ and $b_{\text{\hspace{1pt}main}}^{\text{\hspace{1pt}parax}}$ can directly be estimated via ray tracing, we can derive
\begin{equation}
b_{\text{\hspace{1pt}main}}^{\text{\hspace{1pt}BV}} \approx b_{\text{\hspace{1pt}main}}^{\text{\hspace{1pt}parax}}+\frac{0.531}{1.5}\cdot(b_{\text{\hspace{1pt}main}}^{\text{\hspace{1pt}parax}} - b_{\text{\hspace{1pt}main}}^{\text{\hspace{1pt}blur}})
\end{equation}
and choose $b_{\text{\hspace{1pt}main}}^{\text{\hspace{1pt}BV}}$ to be our final focus point.
\subsection{Magnification}\label{section_measure_m}
\begin{figure}[h]
	\centering
	
	\includegraphics[width=0.7\linewidth]{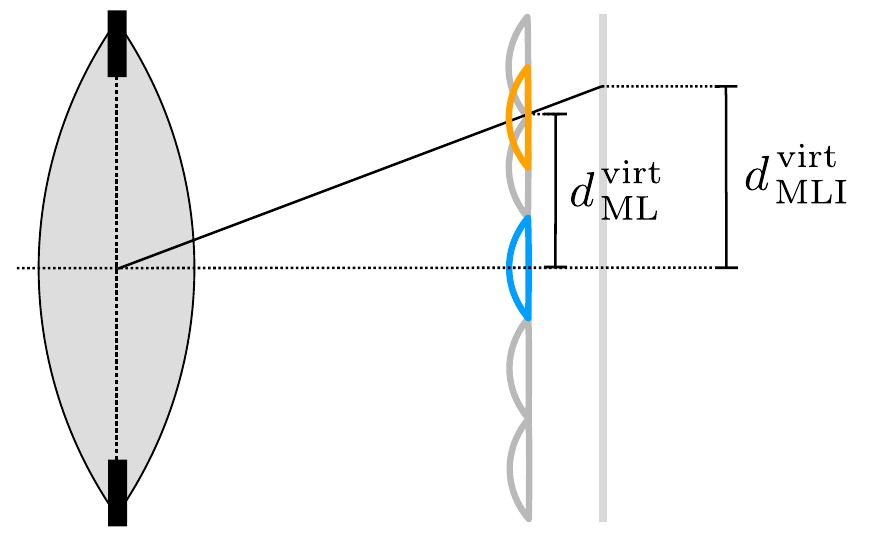}
	\caption{Magnification measurement based on the microlens and MLI centers of the center microlens in blue and a virtual microlens, marked in orange.}
	\label{fig_m}
\end{figure}

For a given plenoptic camera setup, the magnification $m$ between the MLA and sensor can directly be calculated via $m=\frac{d^{\text{\hspace{1pt}virt}}_{\text{\hspace{1pt}MLI}}}{d^{\text{\hspace{1pt}virt}}_{\text{\hspace{1pt}ML}}}$ with $d^{\text{\hspace{1pt}virt}}_{\text{\hspace{1pt}ML}}$ and $d^{\text{\hspace{1pt}virt}}_{\text{\hspace{1pt}MLI}}$
denoting the distances between two microlens centers and the respective MLI centers. To this end a single ray with non-zero angle to the optical axis is traced from the center of the main lens aperture through the remaining main lens and MLA onto the sensor. To get a more accurate result for a slightly defocused setup, it can additionally be assumed, that the ray exactly passes a ML center as indicated by the virtual lens in Fig. \ref{fig_m}.
\subsection{Size of the Visible MLI}\label{section_measure_MLI}
The pixel response on a real sensor does not only depend on the wavelength of the incoming light, but also on its incident angle \cite{hirigoyen2008fdtd}. Accordingly, an angular threshold $\alpha$ can be found, so that the contribution of incoming light with a larger angle to the sensor normal can be assumed to be negligible. For a specific sensor position and microlens in a  given plenoptic camera we can now search for the minimum and maximum incident angles $\alpha_{\text{min}}, \alpha_{\text{max}}\in[-\alpha,\alpha]$ of light rays which are not blocked by the camera housing or aperture. This can be realized by stepping through the interval $[-\alpha,\alpha]$ and tracing rays with the respective angle from the pixel through the currently investigated microlens and the main lens. If such a ray is not blocked or reflected within the camera, the pixel is considered to receive light through that microlens. Starting from the MLI center, a linear search for the first pixel, which does not receive any light through the current microlens, is performed. The maximum distance of a light receiving pixel to the MLI center then defines half of the visible MLI size.
\subsection{Disparity}\label{section_measure_disp}
In order to calculate the disparity of a scene point's images in neighboring microlenses, one can trace a dense bundle of rays from this scene point through the main lens and MLA onto the sensor. Depending on the focus properties of the current setup, this will result in multiple clusters on the sensor. Since we will primarily use the disparity calculation for scene points on the optical axis and within the cameras DoF, the clusters close to the optical axis can be assumed to be sufficiently small to simply use the means of the clusters for the disparity calculation. Given two cluster mean points $\mu_1$ and $\mu_2$ in neighboring MLIs, the disparity coefficient $\gamma$ is given by 
\begin{equation}
\gamma = \frac{||\mu_2-\mu_1||-d_\text{\hspace{1pt}MLI}}{d_\text{\hspace{1pt}MLI}}
\end{equation}
with $d_\text{\hspace{1pt}MLI} =  m\cdot d_\text{\hspace{1pt}ML}$ describing the distance between the MLI centers. 
\subsection{DoF Interval}\label{section_measure_DoF}
First of all, note, that instead of using the size of the area in focus, we define the DoF as interval $[\delta_{\text{\hspace{1pt}min}},\delta_{\text{\hspace{1pt}max}}]\subset\mathbb{R}$ with the boundaries describing the minimum and maximum distance of a scene point from the camera to be in focus.
For well designed conventional cameras, this DoF can be assumed to be similar across all pixel positions on a sensor. Thus the analysis of the DoF is usually limited to paraxial approximations for the sensor center. In a plenoptic camera, the microlenses introduce an additional optical component influencing the DoF of a pixel. In order to estimate this DoF for an arbitrary pixel $x$ in a given camera setup, first the part of the scene that is visible from this pixel is determined by tracing a bundle of rays from $x$ into the scene. The mean ray $\overline{r}$ of the resulting bundle of scene rays is then used as source for scene points for which the blur radius on the sensor is calculated in order to determine, whether these points are in focus. More specifically, all rays traced from $x$ into the scene are intersected with $\overline{r}$ and the average intersection point is used as starting point for the reverse tracing back into the camera. Similar to the calculation of the size of the visible MLI in section \ref{section_measure_MLI}, we iteratively move along $\overline{r}$ in both directions from this starting point and in every step trace a bundle of rays from the currently investigated mean ray point onto the sensor. This procedure is stopped for a direction as soon as the blur radius of a mean ray point's image surpasses the size of a sensor pixel. Hereby, the diameter of the cluster of sensor hits around the investigated pixel $x$ is used for the blur radius calculation. The distances of the closest and farthest points on $\overline{r}$, which are in focus, to the camera define the pixel's DoF $[\delta_{\text{\hspace{1pt}min}}^{\hspace{1pt}x},\delta_{\text{\hspace{1pt}max}}^{\hspace{1pt}x}]$.\\
For the camera parameter optimization regarding a preset DoF, a single DoF range describing the camera's focus performance is required instead of a multitude of DoF intervals depending on the sensor pixel position. Since the variance in the DoF within a microlens is usually larger than the variance across the sensor, the DoF intervals of a single MLI's pixels are used to define the DoF range for the given plenoptic camera setup via
\begin{equation}
\text{DoF} = \left[\max_{x\in \text{MLI}}\{\delta_{\text{\hspace{1pt}min}}^{\hspace{1pt}x}\},\min_{x\in \text{MLI}}\{\delta_{\text{\hspace{1pt}max}}^{\hspace{1pt}x}\}\right].
\end{equation}

\section{FPC Parameter Optimization}\label{section_method}
For FPCs there are two possible configurations depending on the positioning of the MLA and sensor with respect to the main lens focus plane. In a Galilean setup, both components are placed between the main lens and the focus plane, whereas in a Keplerian setup they are located behind the focus plane \cite{georgiev2009dofinplenopticcams}. For the sake of visual simplicity we will only discuss the Keplerian setup as shown in Fig. \ref{fig_thinlens}.

\subsection{Thin Lens Model Constraints}\label{section_thinlens}
\begin{figure}[h]
	\centering
	\includegraphics[width=0.95\linewidth]{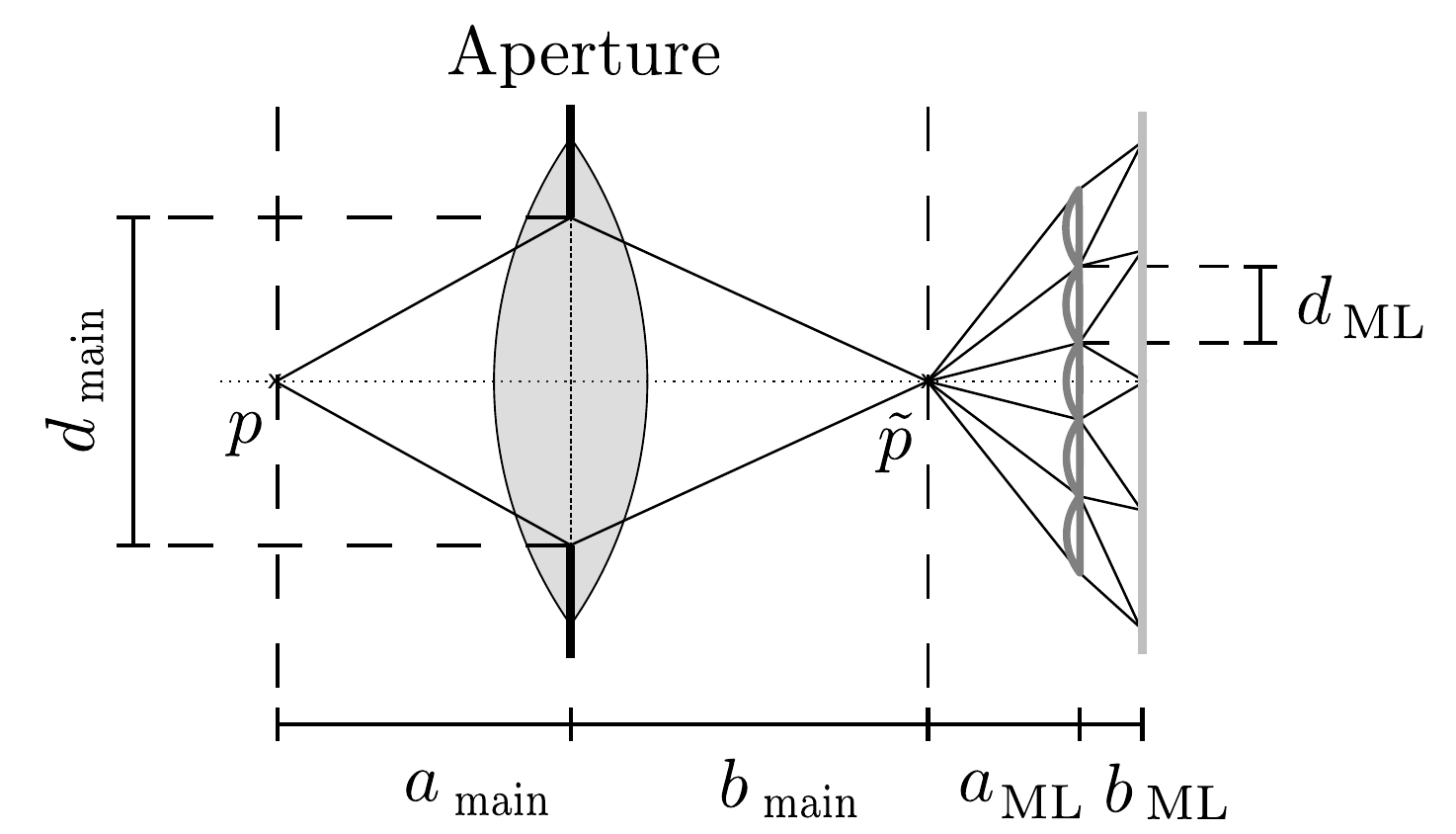}
	\caption{Parameter overview based on the thin lens model of an FPC. The main lens image $\tilde{p}$ of the scene point $p$ is perfectly in focus of the microlenses.}
	\label{fig_thinlens}
\end{figure}
As described before, we will assume a fixed main lens with variable aperture size $d_{\text{\hspace{1pt}main}}$ and are mainly concerned with the focal length $f_{\text{\hspace{1pt}ML}}$ and pitch $d_{\text{\hspace{1pt}ML}}$ of the microlenses and the distances between main lens and MLA as well as between MLA and sensor. In this basic setup, we further assume a scene point $p$ on the optical axis, which should be brought into focus and also have a predefined disparity for this point's sensor image of $disp=\gamma\cdot d_\text{\hspace{1pt}MLI}$ with $\gamma\in(0,1]$ in neighboring MLIs. According to the thin lens equation we get
\begin{equation}\label{eq_mainlens}
\frac{1}{f_{\text{\hspace{1pt}main}}}=\frac{1}{a_{\text{\hspace{1pt}main}}} + \frac{1}{b_{\text{\hspace{1pt}main}}}
\end{equation}
to calculate the position of the virtual image $\tilde{p}$ of $p$ and
\begin{flalign}\label{eq_constraint_1}
\text{\textbf{Constraint 1:}}\quad\frac{1}{f_{\text{\hspace{1pt}ML}}}=\frac{1}{a_{\text{\hspace{1pt}ML}}} + \frac{1}{b_{\text{\hspace{1pt}ML}}}. &&
\end{flalign}
as first constraint for the unknown parameters. A further constraint is given by the f-number matching between main lens and microlenses as described in \cite{perwass2012single} for the purpose of optimal sensor space usage. In our notation (compare Fig. \ref{fig_thinlens}) this reads as
\begin{flalign}\label{eq_constraint2}
\text{\textbf{Constraint 2:}}\quad\frac{b_{\text{\hspace{1pt}ML}}}{d_{\text{\hspace{1pt}ML}}} = \frac{b_{\text{\hspace{1pt}main}} + a_{\text{\hspace{1pt}ML}}}{d_{\text{\hspace{1pt}main}}}.&&
\end{flalign}
The magnification of the image between the MLA position and the sensor position, which can also be regarded as scaling factor for the distances $d_{\text{\hspace{1pt}ML}}$ and $d_\text{\hspace{1pt}MLI}$ between ML centers and MLI centers, is given by the method of similar triangles via
\begin{equation}\label{eq_m}
m = \frac{d_\text{\hspace{1pt}MLI}}{d_{\text{\hspace{1pt}ML}}} =\frac{b_{\text{\hspace{1pt}main}}+a_{\text{\hspace{1pt}ML}}+b_{\text{\hspace{1pt}ML}}}{b_{\text{\hspace{1pt}main}}+a_{\text{\hspace{1pt}ML}}}.
\end{equation}
Analogously we get
\begin{equation}\label{eq_disp_1}
\frac{disp + d_\text{\hspace{1pt}MLI}}{a_{\text{\hspace{1pt}ML}} + b_{\text{\hspace{1pt}ML}}} = \frac{d_{\text{\hspace{1pt}ML}}}{a_{\text{\hspace{1pt}ML}}}
\end{equation}
and thus using $disp=\gamma\cdot d_\text{\hspace{1pt}MLI}=\gamma\cdot m\cdot d_{\text{\hspace{1pt}ML}}$ we get our last constraint
\begin{flalign}\label{eq_constraint3}
\text{\textbf{Constraint 3:}}\quad\frac{(1+\gamma)\cdot m\cdot d_{\text{\hspace{1pt}ML}}}{a_{\text{\hspace{1pt}ML}} + b_{\text{\hspace{1pt}ML}}} = \frac{d_{\text{\hspace{1pt}ML}}}{a_{\text{\hspace{1pt}ML}}}.&&
\end{flalign}
At this point, after applying equation \ref{eq_mainlens} to calculate $b_{\text{\hspace{1pt}main}}$, we have five unknown parameters, $d_{\text{\hspace{1pt}main}}$, $f_{\text{\hspace{1pt}ML}}$, $a_{\text{\hspace{1pt}ML}}$, $b_{\text{\hspace{1pt}ML}}$ and $d_{\text{\hspace{1pt}ML}}$, but only three constraints. However, since the first constraint ensures, that the camera is focused on $p$, the second constraint results in a fully used sensor area, and the last constraint guarantees our desired disparity values for the sensor image of $p$, all of our current requirements are met and we can freely choose two of the parameters.\\
When building one's own plenoptic camera the placement of the MLA and sensor as well as the aperture size of the main lens are usually adjustable while the MLA parameters are often given due to the use of off-the-shelf Hartmann-Shack lens arrays. Accordingly, we will now describe the calculation of the parameters $d_{\text{\hspace{1pt}main}}$, $a_{\text{\hspace{1pt}ML}}$ and $b_{\text{\hspace{1pt}ML}}$ for given MLA parameters $f_{\text{\hspace{1pt}ML}}$ and $d_{\text{\hspace{1pt}ML}}$. We start with solving the last constraint for $b_{\text{\hspace{1pt}ML}}$ resulting in
\begin{equation}\label{eq_sensor}
b_{\text{\hspace{1pt}ML}} = \frac{\gamma\cdot a_{\text{\hspace{1pt}ML}}\cdot (b_{\text{\hspace{1pt}main}} + a_{\text{\hspace{1pt}ML}})}{(b_{\text{\hspace{1pt}main}} + a_{\text{\hspace{1pt}ML}}) - (1+\gamma)\cdot a_{\text{\hspace{1pt}ML}}}.
\end{equation}
Plugging this into equation \ref{eq_constraint_1} and discarding the negative solution of the resulting quadratic equation results in
\begin{equation}\label{eq_tildep}
a_{\text{\hspace{1pt}ML}} = -\frac{b_{\text{\hspace{1pt}main}}}{2} + \sqrt{f_{\text{\hspace{1pt}ML}}\cdot b_{\text{\hspace{1pt}main}}\cdot\frac{1+\gamma}{\gamma} + \frac{b_{\text{\hspace{1pt}main}}^{\hspace{1pt}2}}{4}}.
\end{equation}
Now we can calculate $a_{\text{\hspace{1pt}ML}}$ and use that value to calculate $b_{\text{\hspace{1pt}ML}}$ according to equation \ref{eq_sensor}. Finally, with the second constraint, we can also calculate the main lens aperture via
\begin{equation}\label{eq_aperture}
d_{\text{\hspace{1pt}main}}=\frac{d_{\text{\hspace{1pt}ML}}\cdot(b_{\text{\hspace{1pt}main}}+a_{\text{\hspace{1pt}ML}})}{b_{\text{\hspace{1pt}ML}}}.
\end{equation}

\subsection{Thick Lens Modifications}\label{section_thicklens}
As the analysis will show, the configurations resulting from the thin lens approximations are usually not functioning in the sense that they are neither appropriately focused nor avoid heavy MLI overlapping. The reason for this is that a realistic main lens does not behave like a thin lens. However, by modifying only two aspects of the previous calculations, the estimation of $b_{\text{\hspace{1pt}main}}$ and $d_{\text{\hspace{1pt}main}}$, the results can be significantly improved. 
\begin{figure}[h]
	\centering
	\includegraphics[width=1\linewidth]{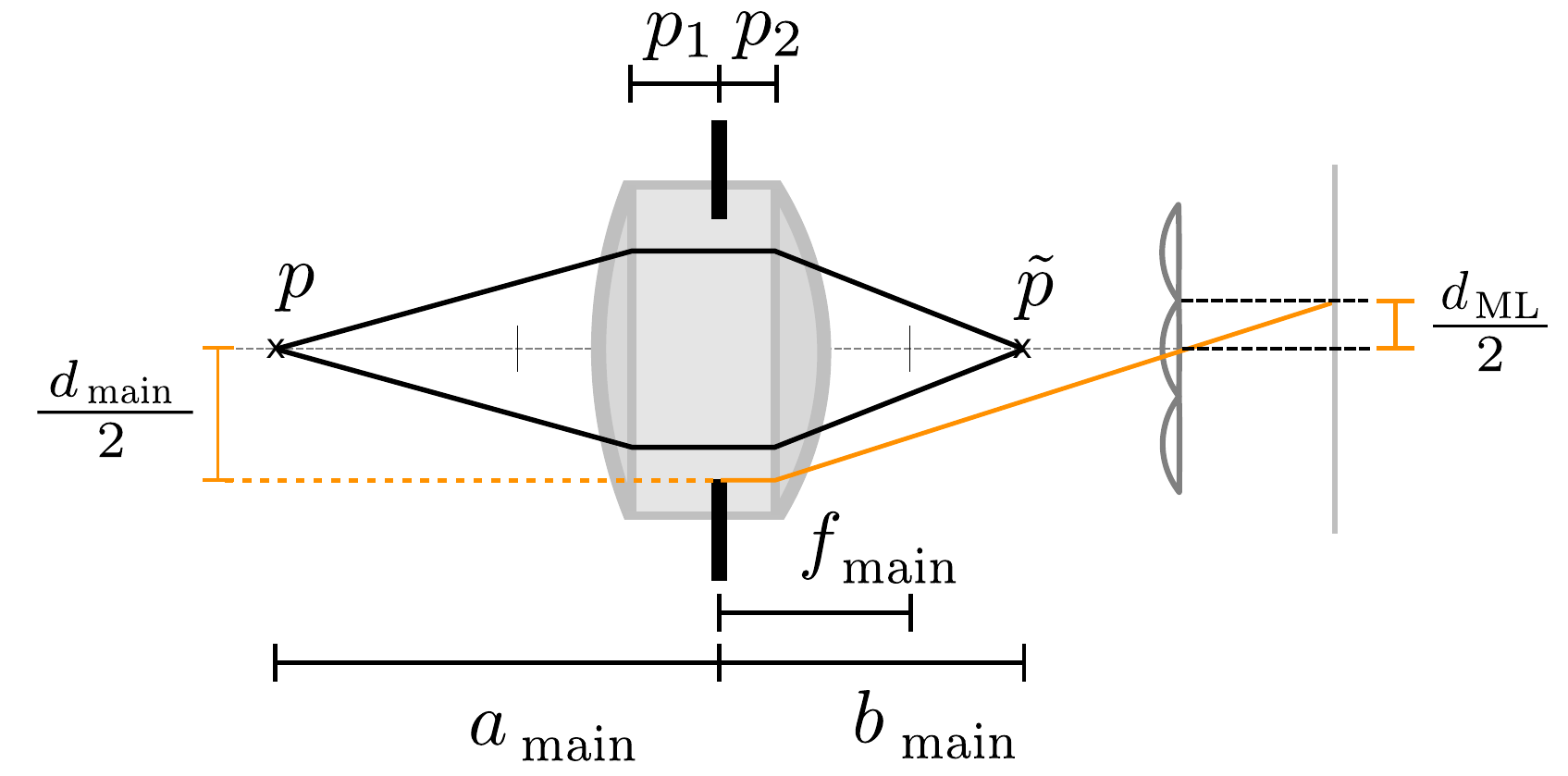}
	\caption{Thick lens model for the main lens. The orange components describe the aperture size calculation.}
	\label{fig_thicklens}
\end{figure}\\
First, equation \ref{eq_mainlens} is replaced by the thick lens equation,
\begin{equation}
\frac{1}{f_{\text{\hspace{1pt}main}}-p_2} = \frac{1}{a_{\text{\hspace{1pt}main}}-p_1} + \frac{1}{b_{\text{\hspace{1pt}main}}-p_2}
\end{equation}
whereby $p_1$ and $p_2$ describe the axial positions of the two principal planes as shown in Fig. \ref{fig_thicklens}. These can be estimated by tracing a single ray from each side of the main lens, parallel to the optical axis, through the main lens and calculating the intersection of the respective resulting and incoming rays. The second modification is a replacement of equation \ref{eq_constraint2} by a simple ray tracing operation. Starting at $(b_{\text{\hspace{1pt}main}}+a_{\text{\hspace{1pt}ML}}+b_{\text{\hspace{1pt}ML}}, 0.5\cdot d_{\text{\hspace{1pt}ML}})$, a ray is traced through the center of the MLA, $(b_{\text{\hspace{1pt}main}}+a_{\text{\hspace{1pt}ML}}, 0)$ into the main lens. Its distance from the optical axis at the aperture position is then used as semi aperture size $d_{\text{\hspace{1pt}main}}/2$. This procedure, visualized in Fig. \ref{fig_thicklens}, effectively replaces the final calculation \ref{eq_aperture} of the last section.\\
While the now enhanced calculations already provide a good starting point for a plenoptic camera design, the parameters still require manual fine tuning due to the approximative nature of the models. The following section aims at minimizing this problem by replacing several of the steps above with ray tracing measurements as presented in the previous section \ref{section_rtmeasure}.

\subsection{Ray Tracing-Based Refinement}\label{section_refinement}
As with the thin lens-based model we start with calculating $b_{\text{\hspace{1pt}main}}$, but use the ray tracing-based procedure of section \ref{section_measure_focus} by tracing rays from $p$ through the main lens and setting $b_{\text{\hspace{1pt}main}}:=b_{\text{\hspace{1pt}main}}^{\text{\hspace{1pt}BV}}$. Since the aperture size $d_{\text{\hspace{1pt}main}}$ is needed during these calculations for determining the position of minimum blur, it is initialized with the value calculated with the thick lens model.\\
With $b_{\text{\hspace{1pt}main}}$ given, the equations \ref{eq_tildep} and \ref{eq_sensor} are then used to calculate initial values for $a_{\text{\hspace{1pt}ML}}$ and $b_{\text{\hspace{1pt}ML}}$ whereby the MLA parameters $f_{\text{\hspace{1pt}ML}}$ and $d_{\text{\hspace{1pt}ML}}$ are assumed to be predefined as in the previous setup. Since the aperture value $d_{\text{\hspace{1pt}main}}$ is already approximately correct and mainly affects the size of the visible MLIs, the main goals are now to focus the camera to the point $p$ and achieve the desired disparity coefficient $\gamma$, both by slightly modifying the MLA and sensor position, i.e. the values $a_{\text{\hspace{1pt}ML}}$ and $b_{\text{\hspace{1pt}ML}}$. In order to do this systematically, we analyze the changes of $\gamma$ and the MLA's focus resulting from modifications of these parameters. First, we solve equation \ref{eq_constraint3} for $\gamma$ and calculate the partial derivatives, which are given by
\begin{align}\nonumber
&\frac{\partial\hspace{1pt} \gamma}{\partial\hspace{1pt} a_{\text{\hspace{1pt}ML}}} = \frac{b_{\text{\hspace{1pt}ML}}\cdot b_{\text{\hspace{1pt}main}}\cdot(b_{\text{\hspace{1pt}ML}}+b_{\text{\hspace{1pt}main}}+2\cdot a_{\text{\hspace{1pt}ML}})}{-d^{\hspace{1pt}2}_{\tilde{p}}\cdot (b_{\text{\hspace{1pt}ML}}+b_{\text{\hspace{1pt}main}}+a_{\text{\hspace{1pt}ML}})^{\hspace{1pt}2}}<0\\\nonumber
&\frac{\partial\hspace{1pt} \gamma}{\partial\hspace{1pt} b_{\text{\hspace{1pt}ML}}} = \frac{b_{\text{\hspace{1pt}main}}\cdot(b_{\text{\hspace{1pt}main}}+a_{\text{\hspace{1pt}ML}})}{a_{\text{\hspace{1pt}ML}}\cdot (b_{\text{\hspace{1pt}ML}}+b_{\text{\hspace{1pt}main}}+a_{\text{\hspace{1pt}ML}})} > 0.
\end{align}
Thus, an increased disparity coefficient can be achieved by either decreasing $a_{\text{\hspace{1pt}ML}}$ or increasing $b_{\text{\hspace{1pt}ML}}$ and analogously, a decreased coefficient requires the respective opposite changes. Similarly from the thin lens equation the necessary modifications for bringing the current focus point of the plenoptic camera onto the sensor can be deduced. If the current focus point is located behind the sensor, either $b_{\text{\hspace{1pt}ML}}$ can be increased to move the sensor to that focus point, or $a_{\text{\hspace{1pt}ML}}$ can be increased to move the focus point closer to the sensor. Again, the opposite case requires the inverse actions. With $\tilde{\gamma}$ denoting the current disparity coefficient calculated as in section \ref{section_measure_disp} and $\tilde{b}_{\text{\hspace{1pt}ML}}$ describing the current distance between the MLA and the focused image of $\tilde{p}$, four cases can be identified
\begin{itemize}\itemsep0em
	\item[1.] $\gamma < \tilde{\gamma}$ and $b_{\text{\hspace{1pt}ML}} < \tilde{b}_{\text{ML}}$: Increase $a_{\text{\hspace{1pt}ML}}$
	\item[2.] $\gamma < \tilde{\gamma}$ and $b_{\text{\hspace{1pt}ML}} > \tilde{b}_{\text{\hspace{1pt}ML}}$: Decrease $b_{\text{\hspace{1pt}ML}}$
	\item[3.] $\gamma > \tilde{\gamma}$ and $b_{\text{\hspace{1pt}ML}} < \tilde{b}_{\text{\hspace{1pt}ML}}$: Increase $b_{\text{\hspace{1pt}ML}}$
	\item[4.] $\gamma > \tilde{\gamma}$ and $b_{\text{\hspace{1pt}ML}} > \tilde{b}_{\text{\hspace{1pt}ML}}$: Decrease $a_{\text{\hspace{1pt}ML}}$
\end{itemize}
These observations can be used to achieve the desired $\gamma$ while keeping the focus point close to the sensor by iteratively performing one of the four actions and recalculating $\tilde{\gamma}$ and $\tilde{b}_{\text{\hspace{1pt}ML}}$ until these are within certain thresholds of the desired values.\\
Finally, the aperture $d_{\text{\hspace{1pt}main}}$ will be adjusted to prevent overlapping MLIs and on the other hand still ensure a good sensor coverage. To do this, we first calculate the current magnification factor $m$ via ray tracing from the aperture center as described in the section \ref{section_measure_m}. Then $m\cdot d_{\text{\hspace{1pt}ML}}$ describes the upper bound for the size $d_{\text{\hspace{1pt}vis}}$ of the visible MLI before overlapping occurs. This size, calculated as in section \ref{section_measure_MLI}, is only measured for a single microlens at or close to the center of the MLA since it is constant across the sensor or decreasing in the case of strong vignetting. Now $d_{\text{\hspace{1pt}main}}$ is modified until $d_{\text{\hspace{1pt}vis}} \approx m\cdot d_{\text{\hspace{1pt}ML}}$, i.e. if $d_{\text{\hspace{1pt}vis}} < m\cdot d_{\text{\hspace{1pt}ML}}$, then $d_{\text{\hspace{1pt}main}}$ is increased, and decreased otherwise. While this adjustment of $d_{\text{\hspace{1pt}main}}$ affects the focus calculations for $b_{\text{\hspace{1pt}main}}$ as well as the MLA's focus point and therefore it might seem necessary to repeat the whole process of this section, our experiments showed, that this final adjustment is rather small due to the accurate initial value and further iterations have no significant effect.

\subsection{DoF Matching}\label{section_dof_matching}
Despite the calculations of the previous section already resulting in a functioning plenoptic camera design, this setup is focused onto a single scene point without considering the camera's DoF. However, since one of the main applications of plenoptic cameras is the depth reconstruction, the DoF plays an important role as it determines the section of the scene which can be reconstructed accurately \cite{perwass2012single}. 
In order to match the FPC's DoF interval to a preset range $[\delta_{\text{\hspace{1pt}min}},\delta_{\text{\hspace{1pt}max}}]$, we first note, that this new requirement forces us to include one of the previously fixed variables into the optimization. While the FPC could be focused on any point within the desired range with the previous method, it is not possible to directly modify the DoF range since the main lens aperture is mainly dependent on the microlens diameter $d_{\text{\hspace{1pt}ML}}$. Since the connection between $d_{\text{\hspace{1pt}ML}}$ and the DoF is simple, namely a larger $d_{\text{\hspace{1pt}ML}}$ leads to a larger aperture $d_{\text{\hspace{1pt}main}}$ and therefore to a decreased DoF, we decide to include this parameter into the optimization, which leaves only $f_{\text{\hspace{1pt}ML}}$ fixed. Overall, the DoF matching uses the general procedure of the previous chapter in an iterative manner as summarized in the following algorithm.
\begin{algorithm}[!h]
	\caption{$\quad $ Meta Optimization for DoF Matching}
	\label{algo_dof_matching}
	\begin{itemize}\itemsep0em 
		\item[1.] Set $p$ to be the point on the optical axis with distance $\frac{\delta_{\text{\hspace{1pt}min}}+\delta_{\text{\hspace{1pt}max}}}{2}$ from the camera (center of preset DoF range).
		\item[2.] Optimize parameters as in section \ref{section_refinement} and calculate initial DoF $[\tilde{\delta}_{\text{\hspace{1pt}min}},\tilde{\delta}_{\text{\hspace{1pt}max}}]$ as in section \ref{section_measure_DoF}.
		\item[3.] while ($\max\{|\tilde{\delta}_{\text{\hspace{1pt}min}}-\delta_{\text{\hspace{1pt}min}}|,|\tilde{\delta}_{\text{\hspace{1pt}max}}-\delta_{\text{\hspace{1pt}max}}|\}>threshold$) 
		\begin{itemize}\itemsep0em 
			\item[i.] Modify $d_{\text{\hspace{1pt}ML}}$ and recalculate parameters as in \ref{section_refinement}  and DoF according to \ref{section_measure_DoF} to bring $\tilde{\delta}_{\text{\hspace{1pt}max}}-\tilde{\delta}_{\text{\hspace{1pt}min}}$ closer to $\delta_{\text{\hspace{1pt}max}}-\delta_{\text{\hspace{1pt}min}}$, i.e. match the DoF sizes.
			\item[ii.] Modify $a_{\text{\hspace{1pt}main}}$ and recalculate parameters as in \ref{section_refinement}  and DoF according to \ref{section_measure_DoF} to bring $\frac{\tilde{\delta}_{\text{\hspace{1pt}min}}+\tilde{\delta}_{\text{\hspace{1pt}max}}}{2}$ closer to $\frac{\delta_{\text{\hspace{1pt}min}}+\delta_{\text{\hspace{1pt}max}}}{2}$, i.e. match the DoF centers.
		\end{itemize}
	\end{itemize}
\end{algorithm}

\section{Evaluation}\label{section_evaluation}
To analyze the presented approach, we implemented the thin and thick lens approximations as well as the optimization via 2D ray tracing in Matlab/Octave and generated camera configuration files which can directly be loaded by the Blender add-on presented in \cite{michels2018simulation}. The exact experiments and their results are described in the following sections.

\subsection{Experiments}\label{section_experiments}
To evaluate the refinement in section \ref{section_refinement}, we first note, that  $a_{\text{\hspace{1pt}main}}$ and $\gamma$ are the only preset variables in that procedure which are only observable and not hardware properties. Accordingly, the two major aspects to evaluate for an FPC design based on that procedure are the focusing properties and the resulting disparities for points located at the focus distance. To this end, five different cameras and five different parameter setups per camera were used to generate a total of 25 FPC designs. The set of cameras comprises of a 6-, a 7- and an 8-element Double-Gaussian objective as well as two fisheye objectives, described in \cite[pp. 306,336,347,160,166]{smith2004modern}. The predefined parameters for the FPCs , apart from the given objective, were $a_{\text{\hspace{1pt}main}}$, $f_{\text{\hspace{1pt}ML}}$, $d_{\text{\hspace{1pt}ML}}$ and $\gamma$ as well as the pixel and the sensor size. The remaining parameters were estimated via the thin lens model, the thick lens model and the proposed method. To evaluate the imaging quality, the resulting models were used in Blender \cite{michels2018simulation} to render images of a binary stripe calibration pattern located at the respective focus distances $a_{\text{\hspace{1pt}main}}$. The maximum widths of the single stripes, were manually set for every setup to ensure the comparability of the 25 different setups. The minimum line width, however, was always set to zero and for each FPC design 20 images in this line width range were rendered.\\
Since the refinement in section \ref{section_refinement} is mainly concerned with focusing the camera onto a point located at the optical axis, only the contrast of the rendered MLI at the image center was evaluated. To this end, first every rendered image was normalized with an additionally rendered white image in order to compensate for vignetting effects \cite{yu2004practical}. Then, a rectangular section $I$ of the center MLI without border pixels, was used for the contrast calculation. With $\mu$ describing the average intensity value of the cutout $I$, the contrast $c$ is calculated via
\begin{equation}\label{eq_contrast}
c = \frac{\sum I(x,y)\cdot\mathbbm{1}_{I(x,y)>\mu}}{\sum \mathbbm{1}_{I(x,y)>\mu}} -\frac{\sum I(x,y)\cdot\mathbbm{1}_{I(x,y)\leq\mu}}{\sum \mathbbm{1}_{I(x,y)\leq\mu}}
\end{equation}
whereby $\mathbbm{1}$ is an indicator function, which is $1$ if the condition is met, and $0$ otherwise. In short, the first fraction describes the mean intensity of the bright values, i.e. the values above $\mu$, and the second fraction analogously calculates the mean intensity of the values below $\mu$. In the case of a value range of $[0,1]$ for $I$ and a perfect camera, the first mean would be $1$ and the second one $0$ resulting in the maximum contrast value $c=1$. This contrast measure, which can be regarded as an approximation of the modulation transfer function, was used to analyze, how well the different models are able to focus onto the preset distance $a_{\text{\hspace{1pt}main}}$.\\
Furthermore, the rendered stripe patterns were also used to evaluate whether the FPC models meet the disparity requirement $\gamma$. This was analyzed by measuring the disparities for the stripe edges in the images rendered with low stripe frequencies since these show the best contrast. More specifically, for a measured pixel distance $d_{\text{\hspace{1pt}px}}$ between a stripe edge and its correspondence in the neighboring MLI and a pixel size of $s_{\text{\hspace{1pt}px}}$, the disparity coefficient can be calculated by
\begin{equation}
\tilde{\gamma} = \frac{d_{\text{\hspace{1pt}px}}\cdot s_{\text{\hspace{1pt}px}} - (m\cdot d_{\text{\hspace{1pt}ML}})}{(m\cdot d_{\text{\hspace{1pt}ML}})}
\end{equation}
because $m\cdot d_{\text{\hspace{1pt}ML}}$ describes the size of an MLI and thereby also the distance between neighboring MLI centers.\\
Since the meta optimization in section \ref{section_dof_matching} uses the refinement procedure which can be assessed by the measures above, the only additional property to evaluate is the FPC's DoF and its alignment with the preset DoF. To this end we generated five additional camera setups, one with each of the formerly listed objectives, and set the parameters $f_{\text{\hspace{1pt}ML}}$ and $\gamma$ as well as the desired DoF interval $[\delta_{\text{\hspace{1pt}min}},\delta_{\text{\hspace{1pt}max}}]$. The threshold in algorithm \ref{algo_dof_matching} was set to $\frac{\delta_{\text{\hspace{1pt}min}}}{1000}$ for the lower DoF bound and to $\frac{\delta_{\text{\hspace{1pt}max}}}{1000}$ for the upper DoF bound. For every of the resulting setup 100 images of calibration patterns located in $[0.75\cdot\delta_{\text{\hspace{1pt}min}}, 1.25\cdot\delta_{\text{\hspace{1pt}max}}]$ were rendered and the contrast profiles based on equation \ref{eq_contrast} were used to analyze the FPC's DoF. Due to its scaling invariance a star-shaped binary pattern was used as calibration pattern for these experiments in order to avoid different contrast results caused by the different magnifications at varying object distances.

\subsection{Results and Discussion}\label{section_results}
\begin{figure}[h]
	\centering
	\includegraphics[width=0.912\linewidth]{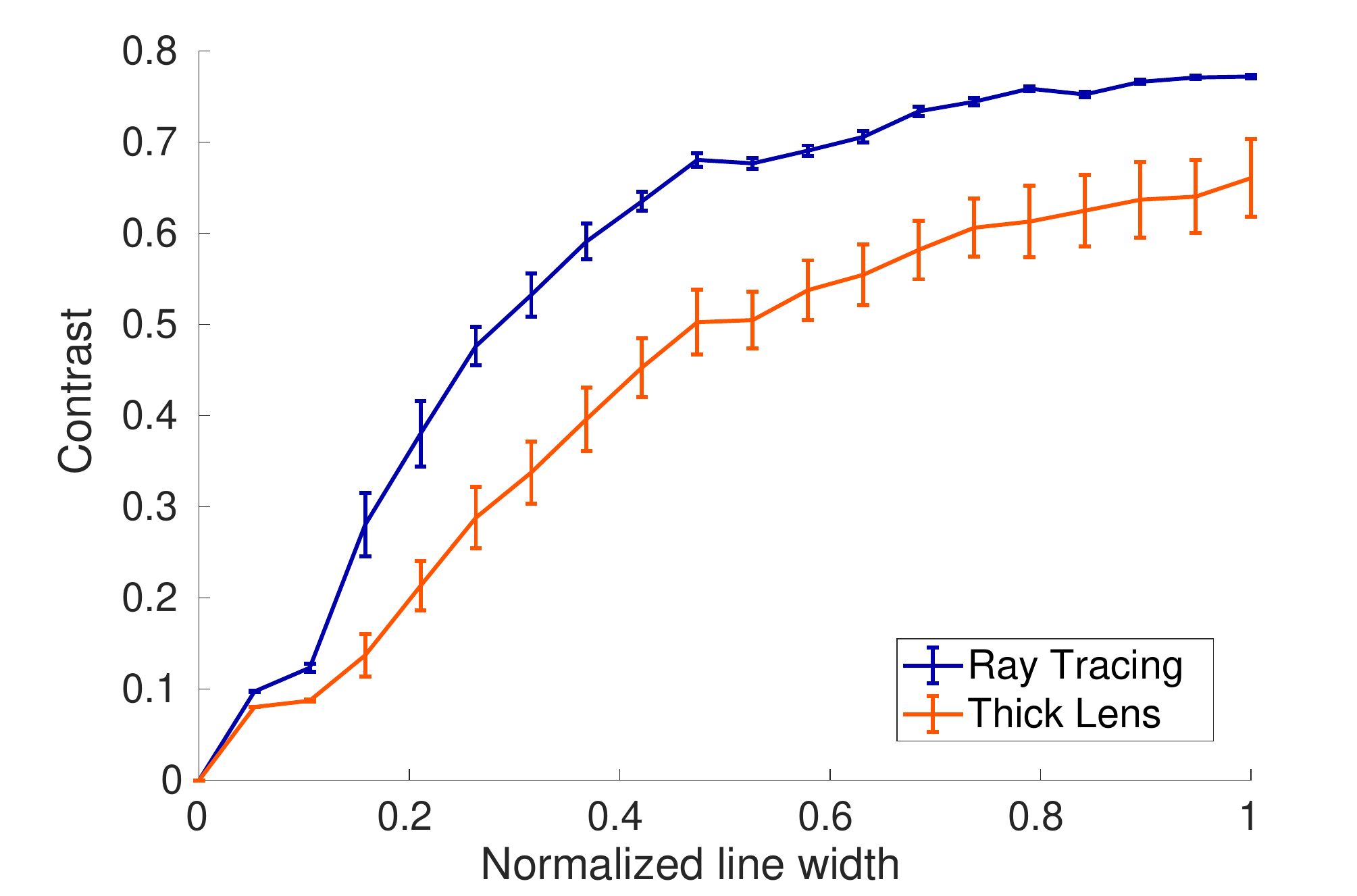}
	\caption{Average contrast and variance of the 25 setups for different line widths.}
	\label{fig_evalcontrast}
\end{figure}

\begin{figure}[t]
	\centering
	\includegraphics[width=1\linewidth]{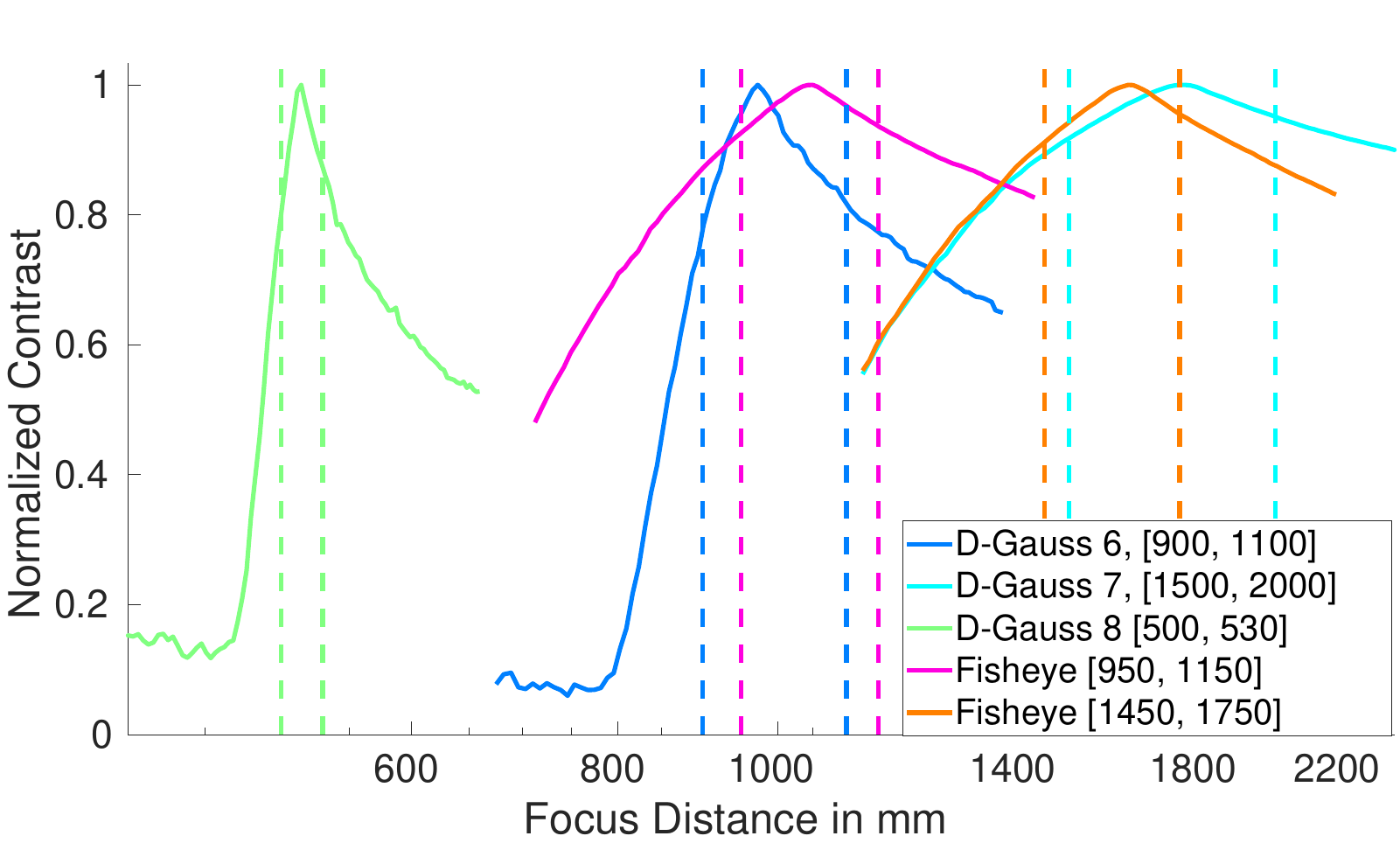}
	\caption{Contrast curves for the five setups. The dotted lines indicate the preset DoF intervals.}
	\label{fig_evaldof}
\end{figure}

In Fig. \ref{fig_evalcontrast} the results of the contrast calculations for the 25 setups are shown in terms of average and variance of contrast. In order to keep the results comparable, all line width ranges were normalized to $[0,1]$, whereby the normalized line width $1$ corresponds to the minimum line frequency of the patterns. 
\begin{figure}[h]
	\centering
	\begin{subfigure}[b]{0.3\linewidth}
		\centering
		\includegraphics[width=\textwidth]{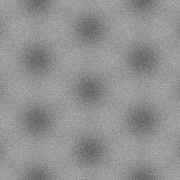}
	\end{subfigure}
	\hfill
	\begin{subfigure}[b]{0.3\linewidth}
		\centering
		\includegraphics[width=\textwidth]{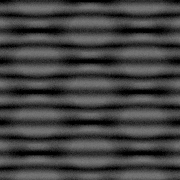}
	\end{subfigure}
	\hfill
	\begin{subfigure}[b]{0.3\linewidth}
		\centering
		\includegraphics[width=\textwidth]{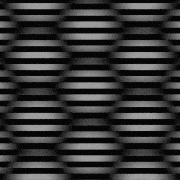}
	\end{subfigure}
	\caption{Exemplary rendering cutouts of the stripe calibration pattern with a thin lens-based model (left), thick lens-based model (middle) and the proposed approach (right). }
	\label{fig_renderings}
\end{figure}
The results for the thin lens-based estimation are not shown due to the simple fact, that these calculations did not produce any usable setups. As exemplarily shown in Fig. \ref{fig_renderings}, the cameras created with the thin lens model were neither focused nor did they avoid heavy MLI overlapping. For the remaining evaluation we will therefore only compare the thick lens-based models with the proposed approach. As shown in Fig. \ref{fig_evalcontrast}, the proposed method outperforms the thick lens model regarding the average contrast as well as the variance. On average, a camera created with our approach shows an increase in contrast by 45\% which is qualitatively supported by the image sections shown in Fig. \ref{fig_renderings}. Moreover, as not visualized here, the worst performing ray tracing-based model still showed a contrast performance comparable to the average thick lens model.\\
Regarding the disparity coefficient, the average difference between the preset $\gamma$ and the $\tilde{\gamma}$ measured as described in the previous section, was $0.0093$ MLI diameters and the maximum deviation was measured to be $0.0269$ MLI diameters. In the case of an MLI diameter of $50\text{\hspace{1pt}px}$ as found in commercially available cameras this translates into an average error of $0.465\text{\hspace{1pt}px}$ and a worst case error of $1.34\text{\hspace{1pt}px}$. A pixel error of this magnitude can be seen as neglectable with respect to the purpose of this constraint, namely keeping the disparities in a range that is well handled by the matching algorithm used in the 3D reconstruction process.\\
Finally, the results of the DoF analysis in Fig. \ref{fig_evaldof} show, that this constraint is also successfully met.
Here, the contrast curves are separately normalized by the respective maximum contrast values for the visual comparison. In all five cases, the focus distance ranges of the highest contrast are within the predefined DoF intervals. Furthermore, in every design a contrast level of at least 75\% of the maximum is given at any point in the preset ranges.\\
In addition to these results, in all experiments, our method successfully prevented MLI overlapping and ensured full sensor coverage, while both, the thin and thick lens estimation failed in several cases.\\
In summary, the constraints, $a_{\text{\hspace{1pt}main}}$, $\gamma$ and the DoF as well as a useful sensor coverage, have been met in all test cases. Moreover, the single-threaded calculation of the first 25 models with our unoptimized Matlab code takes less than 5 seconds on a Ryzen 5 3600, while the meta optimization for the 5 additional models requires 15 seconds and between 5 and 13 iterations until the threshold is met.

\section{Conclusion and Limitations}\label{section_conclusion}
The proposed methods have been shown to significantly outperform the standard approaches usually used to model plenoptic cameras, while simultaneously enabling the definition of additional constraints. However, there are certain limitations to this work. Currently, neither the refinement nor the meta optimization perform sanity checks on the variables. To prevent inexperienced users to enter constraints which simply can not result in a useful camera design, e.g. because the resulting microlens size is of the same order of magnitude as the pixel size, it would be helpful to initially perform rough approximations of the expected parameters. Furthermore, while the meta optimization worked well in our experiments, there is no guarantee for convergence. If the threshold for the DoF in algorithm \ref{algo_dof_matching} is too small, even minimal counteractions of the separate optimization steps could result in the parameters jumping around the optimum, similar to a gradient descent with fixed step size. Nevertheless, when the meta optimization converges, it has been shown to produce useful models, which in the future can speed-up the evaluation of work concerned with the image data of plenoptic cameras.

\section{Acknowledgement}With these last words, we would like to thank Arne 'Plenoptic-Man' Petersen for his invaluable insights into the inner workings of plenoptic cameras.

{\small
	\bibliographystyle{IEEEtran}
\bibliography{bibliography/egbib}

\begin{thebibliography}{10}
\providecommand{\url}[1]{#1}
\csname url@samestyle\endcsname
\providecommand{\newblock}{\relax}
\providecommand{\bibinfo}[2]{#2}
\providecommand{\BIBentrySTDinterwordspacing}{\spaceskip=0pt\relax}
\providecommand{\BIBentryALTinterwordstretchfactor}{4}
\providecommand{\BIBentryALTinterwordspacing}{\spaceskip=\fontdimen2\font plus
\BIBentryALTinterwordstretchfactor\fontdimen3\font minus
  \fontdimen4\font\relax}
\providecommand{\BIBforeignlanguage}[2]{{%
\expandafter\ifx\csname l@#1\endcsname\relax
\typeout{** WARNING: IEEEtran.bst: No hyphenation pattern has been}%
\typeout{** loaded for the language `#1'. Using the pattern for}%
\typeout{** the default language instead.}%
\else
\language=\csname l@#1\endcsname
\fi
#2}}
\providecommand{\BIBdecl}{\relax}
\BIBdecl

\bibitem{lippmann1908integralphoto}
G.~Lippmann, ``Epreuves reversibles, photographies integrales,'' \emph{Academie
  des sciences, 446451}, 1908.

\bibitem{ives1928camera}
H.~E. Ives, ``A camera for making parallax panoramagrams,'' \emph{JOSA},
  vol.~17, no.~6, pp. 435--439, 1928.

\bibitem{adelsonwang1992plenopticcam}
E.~H. Adelson and J.~Y. Wang, ``Single lens stereo with a plenoptic camera,''
  \emph{IEEE transactions on pattern analysis and machine intelligence},
  vol.~14, no.~2, pp. 99--106, 1992.

\bibitem{ng2005lightfieldcamera}
R.~Ng, M.~Levoy, M.~Br{\'e}dif, G.~Duval, M.~Horowitz, and P.~Hanrahan, ``Light
  field photography with a hand-held plenoptic camera,'' \emph{Computer Science
  Technical Report CSTR}, vol.~2, no.~11, pp. 1--11, 2005.

\bibitem{georgiev2006lightfieldcamdesign}
T.~Georgiev and C.~Intwala, ``Light field camera design for integral view
  photography,'' \emph{Adobe Technical Report}, 2006.

\bibitem{perwass2012single}
C.~Perwass and L.~Wietzke, ``Single lens 3d-camera with extended
  depth-of-field,'' in \emph{Human Vision and Electronic Imaging XVII}, vol.
  8291.\hskip 1em plus 0.5em minus 0.4em\relax International Society for Optics
  and Photonics, 2012, p. 829108.

\bibitem{fleischmann2014plenoptic}
O.~Fleischmann and R.~Koch, ``Lens-based depth estimation for multi-focus
  plenoptic cameras,'' in \emph{German Conference on Pattern
  Recognition}.\hskip 1em plus 0.5em minus 0.4em\relax Springer, 2014, pp.
  410--420.

\bibitem{michels2018simulation}
\BIBentryALTinterwordspacing
T.~Michels, A.~Petersen, L.~Palmieri, and R.~Koch, ``Simulation of plenoptic
  cameras,'' in \emph{2018-3DTV-Conference: The True Vision-Capture,
  Transmission and Display of 3D Video (3DTV-CON)}.\hskip 1em plus 0.5em minus
  0.4em\relax IEEE, 2018, pp. 1--4. [Online]. Available:
  \url{https://github.com/Arne-Petersen/Plenoptic-Simulation}
\BIBentrySTDinterwordspacing

\bibitem{kolb1995lensmodelforsimulation}
\BIBentryALTinterwordspacing
C.~Kolb, D.~Mitchell, and P.~Hanrahan, ``A realistic camera model for computer
  graphics,'' in \emph{Proceedings of the 22Nd Annual Conference on Computer
  Graphics and Interactive Techniques}, ser. SIGGRAPH '95.\hskip 1em plus 0.5em
  minus 0.4em\relax New York, NY, USA: ACM, 1995, pp. 317--324. [Online].
  Available: \url{http://doi.acm.org/10.1145/218380.218463}
\BIBentrySTDinterwordspacing

\bibitem{wu2011lensmodelforeffectsimulation}
J.~Wu, C.~Zheng, X.~Hu, and C.~Li, ``An accurate and practical camera lens
  model for rendering realistic lens effects,'' in \emph{Computer-Aided Design
  and Computer Graphics (CAD/Graphics), 2011 12th International Conference
  on}.\hskip 1em plus 0.5em minus 0.4em\relax IEEE, 2011, pp. 63--70.

\bibitem{zheng2017neurolens}
Q.~Zheng and C.~Zheng, ``Neurolens: Data-driven camera lens simulation using
  neural networks,'' in \emph{Computer Graphics Forum}, vol.~36, no.~8.\hskip
  1em plus 0.5em minus 0.4em\relax Wiley Online Library, 2017, pp. 390--401.

\bibitem{zhang2015forwardsimulation}
R.~Zhang, P.~Liu, D.~Liu, and G.~Su, ``Reconstruction of refocusing and
  all-in-focus images based on forward simulation model of plenoptic camera,''
  \emph{Optics Communications}, vol. 357, pp. 1--6, 2015.

\bibitem{liang2015simuWOmainlens}
C.-K. Liang and R.~Ramamoorthi, ``A light transport framework for lenslet light
  field cameras,'' \emph{ACM Transactions on Graphics (TOG)}, vol.~34, no.~2,
  p.~16, 2015.

\bibitem{liu2015numericalsimulation}
B.~Liu, Y.~Yuan, S.~Li, Y.~Shuai, and H.-P. Tan, ``Simulation of light-field
  camera imaging based on ray splitting monte carlo method,'' \emph{Optics
  communications}, vol. 355, pp. 15--26, 2015.

\bibitem{li2017numericalsimulation}
T.-J. Li, S.~Li, Y.~Yuan, Y.-D. Liu, C.-L. Xu, Y.~Shuai, and H.-P. Tan,
  ``Multi-focused microlens array optimization and light field imaging study
  based on monte carlo method,'' \emph{Optics express}, vol.~25, no.~7, pp.
  8274--8287, 2017.

\bibitem{nurnberg2019simulation}
T.~N{\"u}rnberg, M.~Schambach, D.~Uhlig, M.~Heizmann, and F.~P. Le{\'o}n, ``A
  simulation framework for the design and evaluation of computational
  cameras,'' in \emph{Automated Visual Inspection and Machine Vision III}, vol.
  11061.\hskip 1em plus 0.5em minus 0.4em\relax International Society for
  Optics and Photonics, 2019, p. 1106102.

\bibitem{blender}
\BIBentryALTinterwordspacing
Blender. [Online]. Available: \url{https://www.blender.org/}
\BIBentrySTDinterwordspacing

\bibitem{kingslake2009lens}
R.~Kingslake and R.~B. Johnson, \emph{Lens design fundamentals}.\hskip 1em plus
  0.5em minus 0.4em\relax academic press, 2009.

\bibitem{shannon2012applied}
R.~Shannon, \emph{Applied Optics and Optical Engineering V8}.\hskip 1em plus
  0.5em minus 0.4em\relax Elsevier, 2012, vol.~8.

\bibitem{zemax}
\BIBentryALTinterwordspacing
{Zemax OpticStudio}, accessed 26.07.2021. [Online]. Available:
  \url{https://www.zemax.com/pages/opticstudio}
\BIBentrySTDinterwordspacing

\bibitem{hahne2014light}
C.~Hahne, A.~Aggoun, S.~Haxha, V.~Velisavljevic, and J.~C.~J. Fern{\'a}ndez,
  ``Light field geometry of a standard plenoptic camera,'' \emph{Optics
  express}, vol.~22, no.~22, pp. 26\,659--26\,673, 2014.

\bibitem{hahne2019plenoptisign}
C.~Hahne and A.~Aggoun, ``Plenoptisign: an optical design tool for plenoptic
  imaging,'' \emph{SoftwareX}, vol.~10, p. 100259, 2019.

\bibitem{geary2002introduction}
\BIBentryALTinterwordspacing
J.~Geary, \emph{Introduction to Lens Design: With Practical ZEMAX
  Examples}.\hskip 1em plus 0.5em minus 0.4em\relax Willmann-Bell, 2002.
  [Online]. Available: \url{https://books.google.de/books?id=dPWAPQAACAAJ}
\BIBentrySTDinterwordspacing

\bibitem{hirigoyen2008fdtd}
F.~Hirigoyen, A.~Crocherie, J.~M. Vaillant, and Y.~Cazaux, ``Fdtd-based optical
  simulations methodology for cmos image sensor pixels architecture and process
  optimization,'' in \emph{Sensors, Cameras, and Systems for
  Industrial/Scientific Applications IX}, vol. 6816.\hskip 1em plus 0.5em minus
  0.4em\relax International Society for Optics and Photonics, 2008, p. 681609.

\bibitem{georgiev2009dofinplenopticcams}
T.~Georgiev and A.~Lumsdaine, ``Depth of field in plenoptic cameras.'' in
  \emph{Eurographics (Short Papers)}, 2009, pp. 5--8.

\bibitem{smith2004modern}
\BIBentryALTinterwordspacing
W.~Smith, \emph{Modern Lens Design}, ser. McGraw-Hill professional engineering:
  Electronic engineering.\hskip 1em plus 0.5em minus 0.4em\relax McGraw-Hill
  Education, 2004. [Online]. Available:
  \url{https://books.google.de/books?id=6yLJefi\_QagC}
\BIBentrySTDinterwordspacing

\bibitem{yu2004practical}
W.~Yu, ``Practical anti-vignetting methods for digital cameras,'' \emph{IEEE
  Transactions on Consumer Electronics}, vol.~50, no.~4, pp. 975--983, 2004.

\end{thebibliography}
}

\end{document}